\documentclass[sn-nature,Numbered]{sn-jnl}

\usepackage{graphicx}%
\usepackage{multirow}%
\usepackage{amsmath,amssymb,amsfonts}%
\usepackage{amsthm}%
\usepackage{mathrsfs}%
\usepackage[title]{appendix}%
\usepackage{xcolor}%
\usepackage{bm}%
\usepackage{eucal}%
\usepackage{textcomp}%
\usepackage{manyfoot}%
\usepackage{booktabs}%
\usepackage{algorithm2e}
\RestyleAlgo{ruled}
\SetKwComment{Comment}{/* }{ */}
\usepackage{listings}%
\usepackage{caption}
\usepackage{subcaption}
\usepackage{mathtools}  
\usepackage{xfrac}
\usepackage{soul}

\begin{document}

\title[Phase Field Crystal - Field Dislocation Mechanics coupled theory]{Coupling Phase Field Crystal and Field Dislocation Mechanics for a consistent description of dislocation structure and elasticity}

\author[1]{\fnm{Manas V.} \sur{Upadhyay}}\email{manas.upadhyay@polytechnique.edu}

\author[2]{\fnm{Jorge} \sur{Vi\~{n}als}}\email{vinals@umn.edu}

\affil[1]{\orgdiv{Laboratoire de M\'{e}canique des Solides (LMS), \orgname{École Polytechnique, Institut Polytechnique de Paris}, CNRS UMR 7649}, \orgaddress{\street{Route de Saclay}, \postcode{91120}, \city{Palaiseau}, \country{France}}}

\affil[2]{\orgdiv{School of Physics and Astronomy}, \orgname{University of Minnesota}, \orgaddress{\street{116 Church St. SE}, \city{Minneapolis}, \postcode{55455}, \state{Minnesota}, \country{USA}}}

\abstract{This work addresses differences in predicted elastic fields created by dislocations either by the Phase Field Crystal (PFC) model, or by static Field Dislocation Mechanics (FDM). The PFC order parameter describes the topological content of the lattice, but it fails to correctly capture the elastic distortion. In contrast, static FDM correctly captures the latter but requires input about defect cores.
The case of a dislocation dipole in two dimensional, isotropic, elastic medium is studied, and a weak coupling is introduced between the two models. The PFC model produces compact and stable dislocation cores, free of any singularity, i.e., diffuse. The PFC predicted dislocation density field (a measure of the topological defect content) is used as the source (input) for the static FDM problem. This coupling allows a critical analysis of the relative role played by configurational (from PFC) and elastic (from static FDM) fields in the theory, and of the consequences of the lack of elastic relaxation in the diffusive evolution of the PFC order parameter.}

\keywords{dislocations, elasticity, phase field, stresses, distortions}

\maketitle

\section{Introduction}\label{sec:intro}

Research on the motion of individual or a small number of dislocations at the nanoscale is motivated by ongoing advances in diagnostic techniques, for example, high energy X-ray and Bragg coherent diffractive imaging, that have opened the door to detailed nanoscale observation of crystalline defects such as dislocations (line-type defects) \cite{Ulvestad2015, Yau2017, re:shen20, re:wilkin21, re:dresselhaus21} and of emergent, collective mesoscale phenomena \cite{re:suter17,Hanson2018,Bhattacharya2019}. Combining this information with nanoscale theory and computation is key to understanding materials response, performance, and evolution across different length scales. 

Field dislocation mechanics (FDM) \cite{acharya_model_2001, acharya_driving_2003, acharya_constitutive_2004, roy2005finite,acharya_size_2006,re:acharya11,zhang2015single,arora2020finite,re:arora23} and phase field crystal models (PFC) \cite{re:haataja02,re:elder02,re:elder04} are two widely used approaches to study the nanoscale response of crystals containing dislocations \cite{re:salvalaglio19, re:zhao23, re:skogvoll23}. Since fast temporal scales associated with lattice vibration are eliminated in both theories, their computational implementation can efficiently access long time phenomena that are difficult to describe by other methods operating at similar length scales such as molecular dynamics \cite{re:emmerich12,re:momeni18}.

FDM is a continuum theory \cite{acharya_model_2001, acharya_driving_2003, acharya_constitutive_2004, acharya_size_2006,re:acharya11,zhang2015single, zhang2017continuum,re:arora20a,re:arora20b} in which thermal lattice vibration is not considered. 
Rather, the theory focuses on the longer time scale of dislocation motion and chooses as a primary field the polar dislocation density tensor and its associated conservation law of topological charge \cite{re:kosevich79,re:rickman97,re:acharya11,re:perreault16}. 
The theory also includes nonlinear, anisotropic elastic fields, and therefore it can account for topological defects in atomic configurations of nonlinear elastic media. 
The fundamental equations of the theory are directly derived from classical balances of mass and momentum, an elastic incompatibility equation linking the finite elastic distortion to the dislocation density tensor, and a conservation law for Burgers vector in the body. 
An energy dissipation inequality based on the Second Law leads to the equations for the various fields \cite{re:arora20a,re:arora20b}. 
Recently, the theory was extended to account for the role of thermal strain incompatibilities on the conservation of Burgers vector and temperature evolution due to dislocation activity through the first law of thermodynamics \cite{UPADHYAY2020104150,LIMACHAVES2024116763}.
The FDM theory requires some minimal, but essential constitutive input: The mobility law of single defects, and its constitutive nonlinear (and non-monotone) elastic behavior (i.e., nonlinear elasticity and appropriate generalized stacking fault energies \cite{vitek1968intrinsic, vitek1998atomic, vitek2008non}). Nevertheless, the theory does not require slip systems to be known \textit{a priori} in order to define an energy density. Importantly, the dissipation inequality yields the proper nonlinear generalization of the Peach-Koehler force field acting on a dislocation segment.

The PFC model introduces a phase field that describes the temporally coarse grained atomic density of the crystal, and a phenomenologically specified free energy functional determined by lattice symmetry. 
By construction, the minimizer of the free energy is a spatially periodic phase field of the required symmetry \cite{re:elder04}. 
Dissipative evolution of the phase field is assumed as a gradient flow driven by the same free energy functional. 
Prior research has considered both equilibrium properties, such as elastic constants \cite{re:elder07}, as well as many features of inhomogeneous systems such as, for example, dislocation motion \cite{re:berry06,re:skaugen18,re:skaugen18b,re:skogvoll22} and grain boundary structure and motion \cite{re:elder07,re:mellenthin08,re:wu12}. 
Since the configuration of the system is described by a single scalar field, the simultaneous description of mass density and crystalline lattice with the same order parameter constitutes an overspecified problem. 
This has been addressed by the introduction of a dissipative current associated with an independent vacancy diffusion mode in \cite{re:heinonen16} (see, also, \cite{re:acharya22} for further discussion). 
This separation, however, is not sufficient to describe the motion of topological defects in which the difference between mass and lattice velocities has topological content \cite{re:mura63,re:kosevich79,re:acharya22}. 
Such a separation has been recently introduced in the PFC by assuming that the phase field order parameter determines only the transverse/incompatible part of the elastic distortion tensor, with the longitudinal/compatible part remaining free to capture compatible lattice distortion, and satisfy elastic equilibrium \cite{re:skaugen18,re:skaugen18b}. 
However, the evolution equation for the phase field was modified in an \textit{ad hoc} fashion to include an affine distortion derived from the compatible distortion.

A theory that is free of the limitations described above, and that couples FDM and the PFC, was proposed in \cite{re:acharya20}. 
The general approach is motivated by the great success of the Peierls model in elucidating basic dislocation physics, while the theory adds a consistent scheme for regularizing stress fields near dislocation cores. 
It borrows from the PFC model its definition of the order parameter and the associated free energy functional. 
This field naturally incorporates defect cores, which are kept localized and compact by its free energy, and features dissipative motion while preserving topological charge. 
All topological features of defects are maintained by the phase field evolution, which are solely dictated by the symmetry of the lattice (without any consideration of elastic interactions). 
The topological content of the PFC is then coupled to an elastic material described within finite deformation FDM theory. 
Elasticity is coupled back to the evolution of the phase field through consideration of a dissipation inequality for the full model. 

However, a computational implementation of the coupled PFC and FDM model is lacking. 
As a first step, in this work, we propose an algorithm for a weakly coupled version of this theory within a small deformation framework. 
The numerical implementation is used to determine the equilibrium configuration of a dislocation dipole in two dimensions and in an elastically isotropic system. 
The aim of this study is to determine to what extent the PFC and FDM predicted distortions and stresses agree in equilibrium. 

\section{Coupled Phase Field Crystal and static Field Dislocation Mechanics}
\label{sec:methods}

We focus on an analysis of the equilibrium configuration of a dislocation dipole in a two dimensional hexagonal lattice, in the small deformation approximation, starting from a weakly coupled formulation of the theory proposed in \cite{re:acharya20}.

\subsection{Phase Field Crystal}

The PFC model is described by a single scalar order parameter $\psi(\bm{x},t)$, which is a function of position $\bm{x}$ and time $t$. 
The governing phenomenological free energy functional (in dimensionless form) is given by \cite{elder_phase-field_2007}
\begin{equation}
\label{eqn:FSH}
{\CMcal F}_{\rm sh} = \int_S f\left( \psi, \nabla^2 \psi \right) d^{2}x  = \int_S  \left[ \frac{r}{2} \psi^{2} + \frac{1}{4} \psi^{4} + \frac{1}{2} \left[ (1+\nabla^{2})\psi\right]^{2} \right] d^{2}x
\end{equation}
where $r$ is a dimensionless parameter. For $r > 0$, the minimizer is $\psi = 0$. For $r < 0$ there is a continuum band of periodic solutions with either uniaxial or hexagonal symmetry of characteristic wavenumber $q \approx 1$. The amplitude of $\psi$ is proportional to $\sqrt{r}$ for small $r$. The symmetry of the minimizer depends on the value of 
\begin{equation}
\label{eqn:psi0}
\psi_{0} = \frac{1}{\Omega} \int_{\Omega} \psi(\bm{x},t) \ d^{2} x
\end{equation}
where $\Omega$ is the surface area of the system. We focus here on $r < 0$ and values of $\psi_0$ that correspond to a stable hexagonal phase \cite{elder_phase-field_2007}. To lowest order in $r$ (the \lq\lq single mode approximation"), the minimizer is approximately given by,
\begin{equation}
\psi(\mathbf{x}) = \psi_{0} + A_{0} \sum_{n=1}^{2N} e^{i \bm{q}^{n} \cdot \mathbf{x}}
\label{eq:ground_state}
\end{equation}
with $N=3$, $\bm{q}^{1} = \hat{\bm{e}}_2, \bm{q}^{2} = \frac{\sqrt{3} \hat{\bm{e}}_1}{2} - \frac{\hat{\bm{e}}_2}{2}  , \bm{q}^{3} = -\frac{\sqrt{3} \hat{\bm{e}}_1}{2} - \frac{\hat{\bm{e}}_2}{2}  $, $\bm{q}^{4} = -\bm{q}^{1}$, $\bm{q}^{5} = -\bm{q}^{2}$, $\bm{q}^{6} = -\bm{q}^{3}$ and \cite{re:elder04}
\begin{equation}
\label{eqn:A0}
A_{0} = \frac{1}{5} \left( |\psi_{0} | + \frac{1}{3} \sqrt{-15 r - 36 \psi_{0}^{2}} \right) .
\end{equation}

The vectors $\{ \hat{\bm{e}}_{i} \}$ are the Cartesian unit vectors. 
This solution is approximate up to ${\cal O}(r^{1/2})$, with the next order being ${\cal O}(r^{3/2})$. 
However, it is common practice in numerical work involving the PFC model to consider values of $r$ that are ${\cal O}(1)$. 
Higher order terms in $r$ missing in Eq. \eqref{eq:ground_state} don't seem to appreciably modify long wavelength modulation of $\psi$. 
Indeed, the numerical solution for $\psi$ used below does contain the entire sequence of perturbative terms, though the decomposition involving complex amplitudes $A$ in Eqs. \eqref{eq:ground_state} and \eqref{eq:mse} is only approximate. 
In the small deformation limit, the energy of an elastic distortion of Eq. \eqref{eqn:FSH} is isotropic, with the Lam\'{e} coefficients given by $\lambda = \mu = 3 A_0^2$ \cite{re:skaugen18}.

The temporal evolution of $\psi$ is defined as a (nonconserved) $L_2$-gradient flow
\begin{equation}
\label{eqn:L2grflow}
\partial_{t} \psi = - \frac{\delta {\CMcal F}}{\delta \psi} = -r \psi - \psi^{3} - (1+\nabla^{2})^{2} \psi.
\end{equation}

In order to maintain the value of $\psi_0$ fixed during the evolution of $\psi$, we explicitly impose Eq. \eqref{eqn:psi0} as a global constraint. 
This procedure differs from the more conventional approach of rewriting Eq. \eqref{eqn:L2grflow} as a flow locally conserving $\psi$ (\lq\lq conserved order parameter"). 
The variational analysis of \cite{re:acharya20} applies only to the nonconserved case, Eq. \eqref{eqn:L2grflow}. 
It would have to be generalized to the locally conserved case if a conserved order parameter model were to be chosen.
However, in that case the evolution equation for the order parameter contains two additional orders in derivatives of $\psi$, and we see no physical reason to justify the added complexity.

A key tenet of the theory in \cite{re:acharya20} is that the coupling between order parameter configurational distortion and material elastic distortion takes place only over scales that are large compared with the (microscopic) order parameter wavelength. There is no reason to believe that elastic distortion and elastic energies could be meaningfully defined at the microscopic scale of the PFC model. 
As a first step a slowly varying configurational distortion is introduced in a standard way thorough a multiple scale expansion of the order parameter \cite{re:cross93} as follows:
\begin{equation}
\psi(\mathbf{x},t) = \psi_{0} + \sum_{n=1}^{2N} A^{n}(\mathbf{x},t) e^{i \mathbf{q}^{n} \cdot \mathbf{x}},
\label{eq:mse}
\end{equation}
where it is understood that the variation of $A$ on its arguments is slow in a multiple scale expansion sense. From a numerical perspective, the slowly varying amplitudes will be computed below by a local demodulation of the phase field as
\begin{equation}\label{eqn:amp}
A^n (\mathbf{x},t) = \left( \psi(\mathbf{x},t) e^{-i\textbf{q}^{n} \cdot \textbf{x}} \right) * g
\end{equation}
obtained through the convolution with a two-dimensional Gaussian kernel $g$ of width on the order of the lattice parameter $a_0$ of the phase field,
\begin{equation}
g(\mathbf{x}) = \frac{1}{2 \pi a_0^2} \text{e}^{-\frac{|\mathbf{x}|^{2}}{2a_0^2}}.
\end{equation}
With this definition, the configurational distortion tensor of the phase field, $\mathbf{Q}$, is defined as \cite{re:skaugen18}
\begin{equation}\label{eqn:Q}
\bm{Q}(\mathbf{x}) = - \frac{d}{N} \sum_{n = 1}^{N} \mathbf{q}^{n} \otimes \text{Im}\left( \frac{\nabla {A}^n}{A^{n}} \right),
\end{equation}
where $d=2$ is the dimension of the system, $\text{Im}(z)$ is the imaginary part of the complex number $z$, and $N=3$ for a hexagonal lattice in two dimensions \cite{re:skogvoll22}. 
The tensor $\mathbf{Q}$ has the same symmetry properties under rotation as the elastic distortion $\mathbf{U}$ (from FDM), is a functional at each point of the phase field $\psi$, and is defined so as to capture the slowly varying distortion of surfaces of constant $\psi$. 
Therefore this tensor describes a local configurational distortion of the phase field, without endowing it with any elastic properties.

In analogy with linear elasticity, a configurational stress tensor can be defined as 
\begin{equation}
\label{eqn:stressQ}
    \bm{\sigma}^{\text{Q}} = \mathbb{C}:\bm{Q},
\end{equation}
where $\mathbb{C}$ plays the role of a stiffness tensor. 
In the isotropic elastic case it is given as $C_{ijkl} = \lambda \delta_{ij}\delta_{kl} + \mu \left( \delta_{ik}\delta_{jl} + \delta_{il}\delta_{jk}\right)$.

It is also possible to define a configurational stress by directly computing the variation of the phase field free energy to an affine distortion $\Delta {\CMcal F}_{sh} = {\CMcal F}_{sh}[\psi(\mathbf{x}')] - {\CMcal F}_{sh}[\psi(\mathbf{x})] $, with $\mathbf{x}' = \mathbf{x} + \bm{\eta}(\mathbf{x})$ as  \cite{re:skaugen18} 
\begin{equation}
s_{ij}^{\psi} = \frac{\delta \Delta {\CMcal F}_{sh}}{\delta (\partial_{j} \eta_{i})}, \quad \mathbf{s}^{\psi} = \nabla \psi \otimes \nabla \left[ \left( 1 + \nabla^2 \right) \psi \right] - \left[ \left( 1 + \nabla^2 \right) \psi \right] \nabla \nabla \psi + f \mathbb{I},
\end{equation}
where the free energy density $f$ is defined in Eq. (\ref{eqn:FSH}), and $\mathbb{I}$ is the rank two identity tensor. 
A slowly varying configurational stress can be defined by demodulation as $\bm{\sigma}^\psi = \bm{s}^\psi * g$. 
While both $\bm{\sigma}^\psi$ and $\bm{\sigma}^{\text{Q}}$ are linear in the distortion $\mathbf{Q}$, the former is a variation relative to a possibly distorted reference configuration whereas the latter is a variation relative to the undistorted ground state. 
Therefore, they are expected to coincide in weakly distorted regions such as far from dislocation cores, but are expected to differ in the vicinity of defects.

The PFC order parameter supports not only smooth distortions but also isolated, topologically stable, dislocations. 
They have been extensively studied for two dimensional uniaxial and hexagonal patterns of fluid convection \cite{re:siggia81,re:cross93,re:tsimring96,re:boyer02b}. 
For defected crystalline materials, they have been analyzed in \cite{re:skaugen18} in two dimensions, and in \cite{re:skogvoll22} in three dimensions. 
A dislocation in a configuration of $\psi$ at $\bm{x}_{0}$ corresponds to a nonzero $\oint \text{d}\bm{\eta} = \bm{b}$ around a contour containing only that dislocation. 
The corresponding distortion of the phase field leads to $A^{n}(\bm{x}) = |A^n| e^{-i \bm{q}^n \cdot \bm{\eta}+i\phi}$, with $\phi(\bm{x})$ non singular and $\oint \text{d}(\text{arg} \ A^n) = - \bm{q}^n \cdot \bm{b}$. 
Defining a topological charge as $s^n = \frac{1}{2\pi} \bm{q}^n \cdot \bm{b}$, $\oint \text{d}(\text{arg} \ A^n) = - 2\pi s^n$,  so that a dislocation at $\bm{x}_0$ corresponds to vortices in the complex amplitude $A^{n}$, of winding number $s = 0, \pm 1$. 
Note that $\sum_{n=1}^N s^n = \frac{1}{2\pi} \bm{b}_p \cdot \sum_{n=1}^N \bm{q}^n = 0$. 
Therefore, in two dimensions, a phase field dislocation comprises vortices in two of the complex amplitudes ($s^{n} = \pm 1$), whereas the third is non singular ($s^{n} = 0$). 

This nonzero circulation can be transformed into an area integral, leading to the definition of the phase field dislocation density tensor,
\begin{equation}\label{eqn:alphapfc}
\overline{\bm{\alpha}} = \nabla \times \bm{Q}
\end{equation}
Although the tensor $\overline{\bm{\alpha}}$ can be explicitly written in terms of $\psi$ \cite{re:skogvoll22}, in our calculations below we will first determine $\bm{Q}$ from the phase field amplitudes, Eq. \eqref{eqn:Q}, and then $\overline{\bm{\alpha}}$ directly from Eq. \eqref{eqn:alphapfc}.

\subsection{Field Dislocation Mechanics}

In the static setting within a small deformation framework, FDM takes as input a polar dislocation density field $\bm{\alpha}$, and mechanical boundary conditions, and yields the elastic distortion field $\bm{U}$ and the stress field $\bm{\sigma}$. 
The latter is related to the former through the Hooke's law as 
\begin{equation}\label{eqn:sigmaU}
    \bm{\sigma} = \mathbb{C}:\bm{U}
\end{equation}

In a simply connected domain with no dislocations, the elastic distortion field $\bm{U}$ is compatible, such that $\nabla \times \bm{U} =0$ (see Appendix \ref{sec:notation} for notation). 
If a simply connected two-dimensional domain has a distribution of `$p$' point dislocations of Burgers vectors $\bm{b}_p$, the elastic distortion is related to the Burgers vector as $\bm{b}_p = \oint \bm{U} \cdot \text{d}\bm{x}' \neq 0$, where the contour over which the integral is taken encloses only the $p^\text{th}$ dislocation. 
The inequality holds only when the elastic distortion field $\bm{U}$ has an incompatible/transverse component i.e., a component with a non-zero curl. 
The static FDM model starts from a Stokes-Helmholtz type decomposition of a tensor to uniquely express the elastic distortion field $\bm{U}$ into compatible/longitudinal $\bm{U}^\parallel$ (curl free) and incompatible/transverse $\bm{U}^\perp$ (divergence free) components as $\bm{U} = \bm{U}^\parallel + \bm{U}^\perp$, such that $\nabla \times \bm{U}^\parallel = 0 \quad \text{and} \quad \nabla \cdot \bm{U}^\perp = 0$. 
In small deformation, the polar dislocation density field $\bm{\alpha}$ is defined as $\bm{\alpha} := \nabla \times \bm{U} = \nabla \times \bm{U}^\perp$. 
Then, for a given $\bm{\alpha}$ field, $\bm{U}^\perp$ can be obtained by solving the following Poisson-type equation
\begin{equation}\label{eqn:Uperp}
    \nabla^2 \bm{U}^\perp = - \nabla \times \bm{\alpha} .
\end{equation}

The compatible part can be obtained by using the elastic constitutive relationship (\ref{eqn:sigmaU}), the Stokes-Helmholtz decomposition, $\bm{U}^\perp$ obtained from equation (\ref{eqn:Uperp}), and the static equilibrium condition $\nabla \cdot \bm{\sigma} = 0$. One finds,
\begin{equation}\label{eqn:Uparallel}
    \nabla \cdot \left( \mathbb{C}:\bm{U}^{\parallel} \right) = - \nabla \cdot \left( \mathbb{C}:\bm{U}^{\perp} \right) .
\end{equation}
Following the computation of $\bm{U}^\perp$ and $\bm{U}^\parallel$, $\bm{\sigma}$ can be obtained by using Eq. (\ref{eqn:sigmaU}).

An equivalent approach to obtain $\bm{\sigma}$ from the dislocation density tensor in an isotropic medium is \cite{re:dewit60,re:kroner81}
\begin{equation}\label{eqn:sigmaKD}
    \bm{\sigma}^{\prime} = 2\mu \left[ \nabla^2 \bm{\chi} + \frac{m}{m-1} \left(\nabla \nabla - \nabla^2 \mathbb{I} \right) \text{tr}(\bm{\chi}) \right]
\end{equation}
where $m = \frac{2(\lambda+\mu)}{\lambda}$ and $\bm{\chi}$ is a stress function that is related to $\bm{\alpha}$ as
\begin{equation}\label{eqn:chi}
    \nabla^2 \nabla^2 \bm{\chi} = \text{sym} \left(\bm{\alpha} \times \nabla \right)
\end{equation}
where the operator ``sym" extracts the symmetric part of a second order tensor.

\subsection{Weak coupling}

We want to examine the equilibrium configuration of a defected medium comprising a stationary dislocation dipole under periodic boundary conditions. An initial configuration comprising a dislocation dipole is considered in a climb configuration. For values of $r \approx 1$, the order parameter relaxes everywhere according to Eq. (\ref{eqn:L2grflow}), but the location of the dislocation cores remains constant for very long times due to Periels barriers to motion \cite{re:boyer02b}. Therefore, at long times, the field $\psi$ is a constrained minimizer of ${\CMcal F}_{sh}$ given the location and Burgers vectors of the two dislocations. The phase field is used to compute $\overline{\bm{\alpha}}$ (Eq. (\ref{eqn:alphapfc})). Next we set $\bm{\alpha} = \overline{\bm{\alpha}}$ and solve the resulting static FDM problem for an elastic medium under periodic boundary conditions. The restricted assumption that the static FDM and the PFC only couple through the dislocation density tensor implies that the incompatible part $\bm{U}^{\perp}$ is equal to the incompatible part of $\bm{Q}$. However, importantly, the compatible part of $\bm{U}$ i.e., $\bm{U}^{\parallel}$, is not equal to the compatible part of $\bm{Q}$. Determination of $\bm{U}^{\parallel}$ requires the resolution of equation (\ref{eqn:Uparallel}).

\section{Algorithm and numerical implementation}

A uniform spatial grid is introduced in a rectangular domain with $N_1$ points along $x_1$ and  $N_2$ along $x_2$. The grid spacing is $\Delta x_1 = a_0/{7}, \Delta x_2 = \sqrt{3} a_0/12$, with $a_0 = 4 \pi / \sqrt{3}$ as dictated by the hexagonal ground state of the free energy in Eq. \eqref{eqn:FSH}. The lateral size of the domain along $x_1$ and $x_2$ is $w = N_1 \Delta x_1 $ and $h = N_2 \Delta x_2$, respectively (Figure \ref{fig:figure1}a) as we assume periodic boundary conditions along both directions. Given these boundary conditions, Fourier transforms of the variables are introduced. Spatial derivatives are computed by either their spectral representation or by central finite differences, whereas time discretization is performed using a forward Euler scheme. 

The initial condition for the order parameter $\psi$ includes $p=2$ dislocations at $(x_{1,p},x_{2,p})$,
\begin{equation}\label{eqn:psiinit}
    \psi^{t= 0} = \psi_0 + A_0 \sum_{n=1}^{2N} \exp{\left[{i \left( \bm{q}^n \cdot \bm{x} + \sum_p s^{n,p} \tan^{-1} \left( \frac{x_2 - x_{2,p}}{x_1 - x_{1,p}} \right) \right) }\right]}
\end{equation}
with $A_0$ computed using Eq. (\ref{eqn:A0}). 
For the current time step $t+\Delta t$, with all the quantities known at the previous time step $t$, the PFC problem is iterated first according to,
\begin{equation}\label{eqn:forwardeulerpsi}
    \psi^{t+\Delta t} = \left( 1 - \Delta t \ \mathcal{L} \right) \psi^{t} - \Delta t (\psi^{t})^3
\end{equation}
where $\mathcal{L} = \left( (r+1) + 2 \nabla^2 + \nabla^2 \nabla^2 \right)$ is the linear operator of Eq. (\ref{eqn:L2grflow}) acting on $\psi$, $\Delta t$ is the time step, and quantities with superscript $t$ refer to the previous time step. 
The first term on the right hand side of Eq. (\ref{eqn:forwardeulerpsi}), $\psi' = (1 - \Delta t \ \mathcal{L}) \psi^t$ can be rewritten as an approximation of $\mathcal{F}^{-1}\left( e^{-\Delta t  \tilde{\mathcal{L}}} \tilde{\psi}^t \right) \approx \mathcal{F}^{-1}\left[ \left(1 - \Delta t \tilde{\mathcal{L}}\right) \tilde{\psi}^t \right] = (1-\Delta t \ \mathcal{L}) \psi^t$, where  $\tilde{\mathcal{L}} = \left( (r+1) - 2 |\bm{k}|^2 + |\bm{k}|^4 \right)$, and $\bm{k}$ is the wavevector in Fourier space. 
The symbol $\mathcal{F}$ stands for Fourier transform, and a tilde over a variable denotes its Fourier transform. 
For better accuracy, the expression involving the exponential function in Fourier space is used. 
Only wavenumbers $|\bm{k}| \leq 2$ are considered so that the first term in Eq. (\ref{eqn:forwardeulerpsi}) is computed as,
\begin{equation}
     \psi' = \mathcal{F}^{-1}\left( e^{-\Delta t  \tilde{\mathcal{L}}} \tilde{\psi}^t \tilde{\Lambda} \right)
\end{equation}
with
\begin{equation}
     \tilde{\Lambda}  =
    \begin{cases}
      1 & \text{for $|\bm{k}| \leq k_l$}\\
      \frac{k_h - |\bm{k}|}{k_h - k_l} & \text{for $k_l < |\bm{k}| \leq k_h$}\\
      0 & \text{for $|\bm{k}| > k_h$}
    \end{cases}       
\end{equation}
where $k_l = 1.4$ and $k_h = 2$ are user-defined constants. 

Finally, $\psi^{t+\Delta t}$ is computed as
\begin{equation}\label{eqn:finalpsievol}
    \psi^{t+\Delta t} = \psi' - \Delta t (\psi^{t})^3
\end{equation}
Once $\psi^{t+\Delta t}$ has been obtained, it is used to find $A^{n,t+\Delta t}$ using,
\begin{equation}\label{eqn:amp_num}
A^{n,t+\Delta t} = \mathcal{F}^{-1}\left[ \mathcal{F}\left( \psi^{t+\Delta t} e^{-i\textbf{q}^{n} \cdot \bm{x}} \right) \tilde{g} \right],
\end{equation}
which in turn is used to compute $\bm{Q}^{t+\Delta t}$ using the central difference approximation to Eq. (\ref{eqn:Q}). 
Then, $\bm{\sigma}^{\text{Q},t+\Delta t}$ is straightforwardly obtained using Eq. (\ref{eqn:stressQ}). 
The stress $\bm{\sigma}^{\psi,t+\Delta t}$ is obtained by first computing $\bm{s}^{\psi,t+\Delta t}$ and then computing
\begin{equation}\label{eqn:stresspsi}
    \bm{\sigma}^{\psi,t+\Delta t} = \mathcal{F}^{-1} \left( \tilde{\bm{s}}^{\psi,t+\Delta t} \tilde{g}  \right)
\end{equation}
Next, $\overline{\bm{\alpha}}^{t+\Delta t}$ is computed using $\bm{Q}^{t+\Delta t}$ and Eq. (\ref{eqn:alphapfc}). 
The result is assigned to $\bm{\alpha}^{t+\Delta t}$, and the elastic problem under periodic boundary conditions is solved as follows: We obtain $\bm{U}^{\perp}$ from (\ref{eqn:Uperp}) using the Fourier transform as
\begin{equation}\label{eqn:Uperpsol}
    \bm{U}^\perp = \mathcal{F}^{-1}  \left( 
    \begin{cases}
         i \frac{\left( \tilde{\bm{\alpha}} \cdot \mathbf{X} \right) \cdot \bm{k} }{|\bm{k}|^2}  & \text{for } |\bm{k}| \neq 0 \\
        0 & \text{for } |\bm{k}| = 0
    \end{cases} \right) ,
\end{equation}
where $\mathbf{X}$ is defined in Appendix \ref{sec:notation}. Then
\begin{equation}\label{eqn:Uparallel_num}
    \bm{U}^\parallel = \mathcal{F}^{-1} \left[\tilde{\mathbb{G}} : \mathbb{C}: \tilde{\bm{U}}^\perp \right]
\end{equation}
where  $\tilde{\mathbb{G}}$ is the fourth order modified Green's tensor with components in Fourier space given by
\begin{equation}
    \tilde{\mathbb{G}}_{ijkl} = 
    \begin{cases}
        - C_{kmin}^{-1} & \text{for any component of } \bm{k} \text{ equal to } 0 \\
        - \left[ k_m k_n C_{kmin} \right]^{-1} k_l k_j & \text{otherwise}
    \end{cases}
\end{equation} 
The stress $\bm{\sigma}^{t+\Delta t}$ is then straightforwardly obtained from (\ref{eqn:sigmaU}) using $\bm{U}^{t+\Delta t} = \bm{U}^{\parallel,t+\Delta t} + \bm{U}^{\perp,t+\Delta t}$. 

Finally, $\bm{\sigma}^{\prime,t+\Delta t}$ is computed from (\ref{eqn:sigmaKD}) using the Fourier transform,
\begin{equation}\label{eqn:sigmaKD_num}
    \bm{\sigma}'^{,t+\Delta t} = \mathcal{F}^{-1} \left[ 2\mu \left( -|\bm{k}|^2 \tilde{\bm{\chi}}^{t+\Delta t} + \frac{m }{m-1} \text{tr}(\tilde{\bm{\chi}}^{t+\Delta t}) \left( |\bm{k}|^2 \mathbb{I} - \bm{k} \otimes \bm{k} \right)  \right) \right]
\end{equation}
with 
\begin{equation}
    \tilde{\bm{\chi}}^{t+\Delta t} = 
    \begin{cases}
        i \frac{\left( \mathbf{X}\cdot \tilde{\bm{\alpha}}^{t+\Delta t} + \tilde{\bm{\alpha}}^{\text{T},t+\Delta t}\cdot \mathbf{X} \right) \cdot \bm{k}}{|\bm{k}|^4} & \text{for } |\bm{k}| \neq 0 \\
        0 & \text{for } |\bm{k}| = 0
    \end{cases}
\end{equation} 
obtained from Eq. (\ref{eqn:chi}).

Note that the mean values of both $\bm{\sigma}^{t+\Delta t}$ and $\bm{\sigma}^{\prime,t+\Delta t}$ are undefined.  We set their Fourier amplitudes at $|\bm{k}| = 0$ to be zero as appropriate for a system with periodic boundary conditions. In order to facilitate comparison with $\bm{\sigma}^{\text{Q},t+\Delta t}$ and $\bm{\sigma}^{\psi,t+\Delta t}$, we also set the mean values of the configurational stresses to be zero.

The step by step algorithm is summarized in Table \ref{alg}, which has been implemented in Fortran90. The FFTW 3 \cite{FFTW05} C++ library is used to perform forward and inverse Fourier transforms. In Fourier space, derivatives (in real space) are approximated using the central difference approach \cite{BERBENNI20144157,UPADHYAY2016126}.

\begin{algorithm}
\caption{PFC-FDM weak coupling}\label{alg}
\KwData{$\bm{x}, \bm{x}_p, \psi_0, r, s^n_p, \bm{q}^n, dx, dy, N_x, N_y, \Delta t, T, \mathbb{C}, k_l, h_h $ }
\KwResult{$\bm{Q}$, $\bm{U}$, $\bm{\sigma}^{\text{Q}}$, $\bm{\sigma}^{\psi}$, $\bm{\sigma}^{\text{FDM}}$, $\bm{\sigma}^{\text{KD}}$}
$A_0 \gets$ equation (\ref{eqn:A0}) \;
$t \gets 0$ \;
$\psi^{t=0} \gets$ R.H.S. of equation (\ref{eqn:psiinit}) \;
\While{$t < T$}{
    
    $\psi^{t+\Delta t} \gets$ R.H.S. of equation  (\ref{eqn:finalpsievol}) \; $\bm{\sigma}^{\psi,t+\Delta t} \gets$ R.H.S. of equation (\ref{eqn:stresspsi}) \;
    $\bm{\sigma}^{\psi,t+\Delta t} \gets \bm{\sigma}^{\psi,t+\Delta t} - \langle \bm{\sigma}^{\psi,t+\Delta t} \rangle$ \;
    $A^{n,t+\Delta t} \gets$ R.H.S. of equation (\ref{eqn:amp_num}) \;
    $\bm{Q}^{t+\Delta t} \gets$ R.H.S. of equation (\ref{eqn:Q}) \;
    $\bm{\sigma}^{\text{Q},t+\Delta t} \gets$ R.H.S. of equation (\ref{eqn:stressQ}) \;
    $\bm{\sigma}^{\text{Q},t+\Delta t} \gets \bm{\sigma}^{\text{Q},t+\Delta t} - \langle \bm{\sigma}^{\text{Q},t+\Delta t} \rangle$ \;
    $\overline{\bm{\alpha}}^{t+\Delta t} \gets$ R.H.S. of equation (\ref{eqn:alphapfc}) \;
    $\bm{\alpha}^{t+\Delta t} \gets \overline{\bm{\alpha}}^{t+\Delta t}$ \;
    $\bm{U}^{\perp, t+\Delta t} \gets $ R.H.S. of equation (\ref{eqn:Uperpsol}) \;
    $\bm{U}^{\parallel, t+\Delta t} \gets $ R.H.S. of equation (\ref{eqn:Uparallel_num}) \;
    $\bm{\sigma}^{t+\Delta t} \gets $ R.H.S. of equation (\ref{eqn:sigmaU}) \;
    $\bm{\sigma}'^{,\ t+\Delta t} \gets $ R.H.S. of equation (\ref{eqn:sigmaKD_num}) \;
    $\psi^{t} \gets$  $\psi^{t+\Delta t}$ \; 
    $t \gets t+\Delta t$ \; 
    }
\end{algorithm}

\section{Results}\label{sec:results}

\begin{figure*}[ht!]%
\centering
\includegraphics[width=\textwidth]{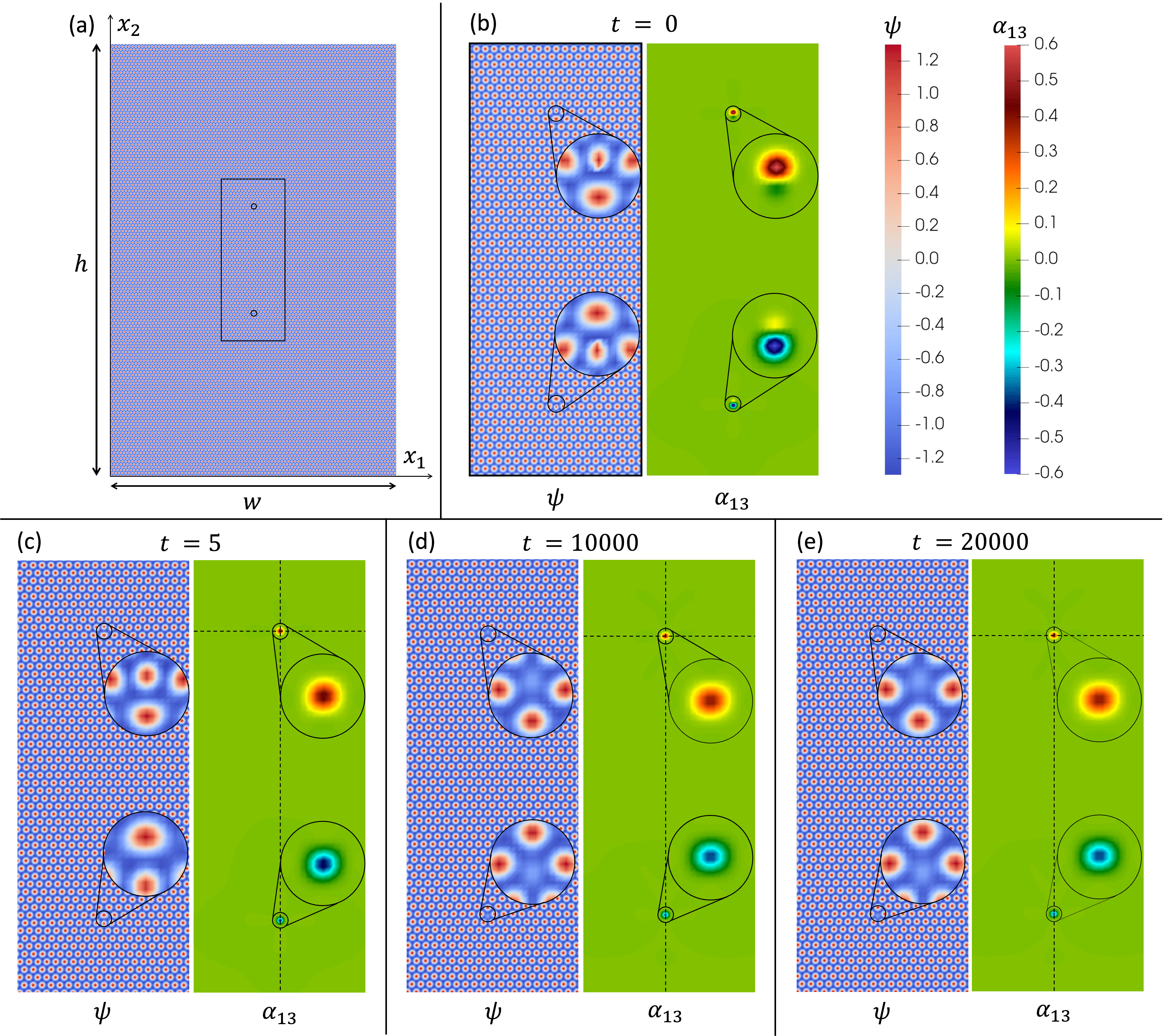}
\caption{Time evolution of the order parameter $\psi$ and dislocation density component $\alpha_{13}$. (a) Initial $\psi$ field in the domain along with the basis and domain dimensions. (b) Close up views (from the highlighted region in (a)) of $\psi$ and $\alpha_{13}$ showing the two dislocations in the domain at $t=0$. The insets zoom into the dislocation core. Snapshots of $\psi$ and $\alpha_{13}$ (in the same zones as those shown in (b)) at (c) $t = 5$, (d) $t = 10000$ and (e) $t = 20000$. All the $\psi$ plots share the same scale. The dotted lines in the $\alpha_{13}$ plots in (c), (d) and (e) are the lines along which the lineplots shown in Figures \ref{fig:stress_vertical} and \ref{fig:stress_horizontal} have been made.}\label{fig:figure1}
\end{figure*}

We have used $N_1 = 602$ points along $x_1$ and  $N_2 = 900$ along $x_2$, and the parameter values $r = -1.2, \psi_{0} = -0.5$, well inside the hexagonal region of the phase diagram of the phase field. 
As initial condition, two dislocations are introduced along the line $y = \sfrac{h}{2}$ according to equation \eqref{eqn:psiinit} with Burgers vectors $\mathbf{b}_{1} = a_{0} (1,0)$ at $\left( x_{1,1}, x_{2,1} \right) = \left( \sfrac{ w}{2}, \sfrac{3 h}{8} \right)$ (top), and  $\mathbf{b}_{2} = a_{0} (-1,0)$ at $\left( x_{1,2}, x_{2,2} \right) = \left( \sfrac{ w}{2}, \sfrac{5 h}{8} \right)$ (bottom), as shown in Figure \ref{fig:figure1}b. 

\begin{figure}[ht!]%
\centering
\includegraphics[width=0.95\textwidth]{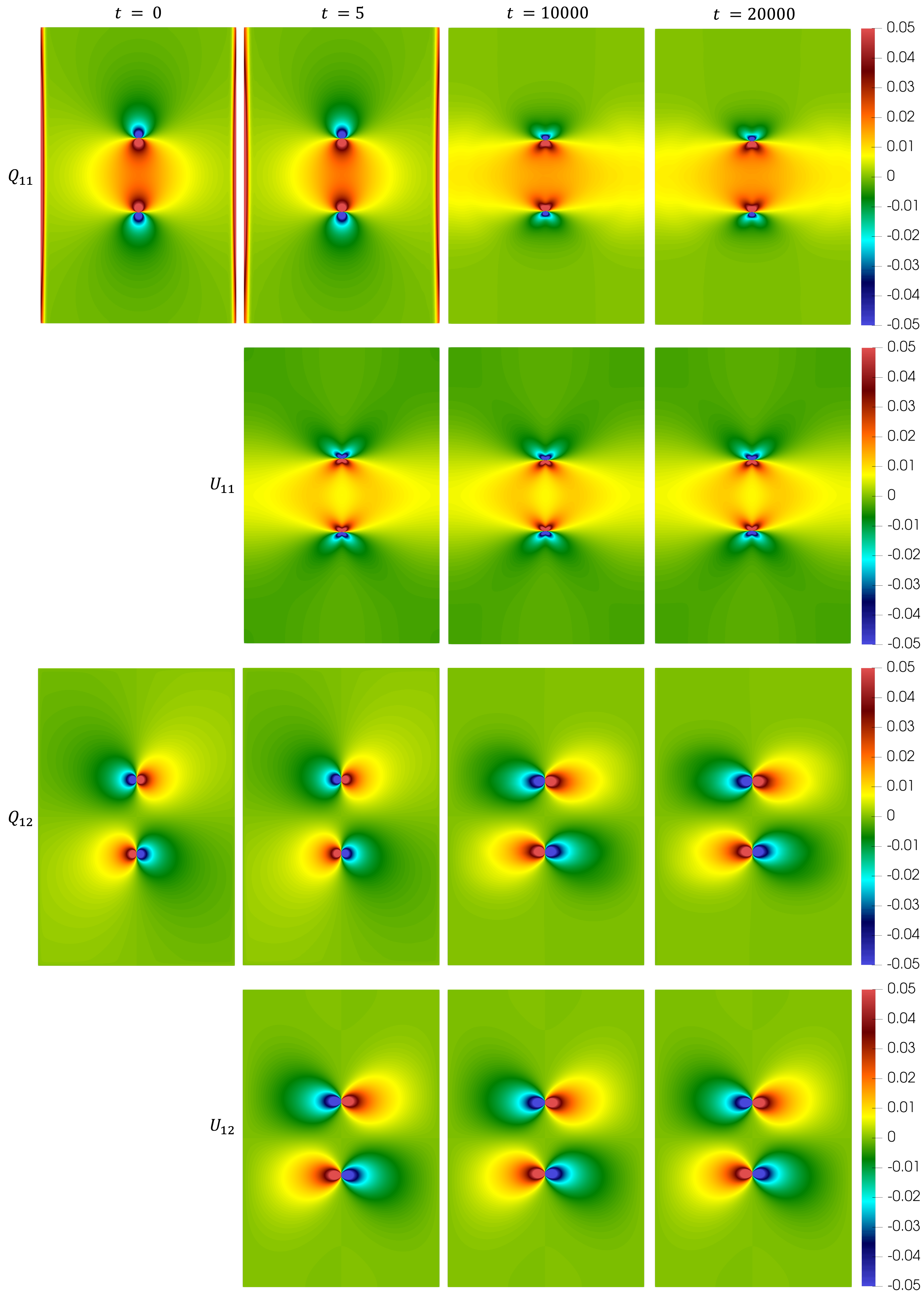}
\caption{Time evolution of $Q_{11}$, $U_{11}$, $Q_{22}$ and $U_{22}$. $U_{11}$ and $U_{22}$ have not been presented at $t = 0$ due to the instability of dislocation cores in the initial state. }\label{fig:UQevolution}
\end{figure}

With this initial configuration, the only non zero component of the polar dislocation density is $\alpha_{13}$. 
The top and bottom dislocations should respectively exhibit $\alpha_{13} > 0$ and $\alpha_{13} < 0$. 
Figure \ref{fig:figure1} shows the temporal evolution of $\psi$ and $\alpha_{13}$ as the system relaxes from the initial condition. 
The order parameter $\psi$ reaches a steady state around $t= 20000$, therefore, only results up to this time are shown. 
However, and as expected from the diffusive nature of the relaxation of the order parameter, $\alpha_{13}$ reaches a time independent distribution much earlier ($t \approx 5$ in the figure). 
As a consequence, the contour lines of $\bm{U}$ do not change or move significantly from this time until the end of the integration (Figure \ref{fig:UQevolution}) despite the fact that $\bm{Q}$ continues to evolve. 
Indeed, after relaxation, the top and bottom dislocations respectively exhibit $\alpha_{13} > 0$ and $\alpha_{13} < 0$. 

Even though $\alpha_{13}$ becomes time independent as early as $t=5$, there are clear qualitative and quantitative differences between the spatial distributions of the corresponding components of $\bm{Q}$ and $\bm{U}$ (Figure \ref{fig:UQevolution}).
These differences are most pronounced in the case of $U_{11}$ and $Q_{11}$. The former shows fourfold symmetry, which is the expected symmetry of this component of the elastic distortion for an edge dislocation with $\alpha_{13}$ as the only non zero component. Meanwhile, $Q_{11}$ shows a twofold geometry, which does not conform with the predictions of the elasto-static theory of dislocations i.e., static FDM. The differences are less pronounced in the case of $Q_{12}$ and $U_{12}$, nevertheless, they do exist, as evidenced in figure \ref{fig:UQevolution}.

\begin{figure*}[ht!]%
\centering
\includegraphics[width=\textwidth]{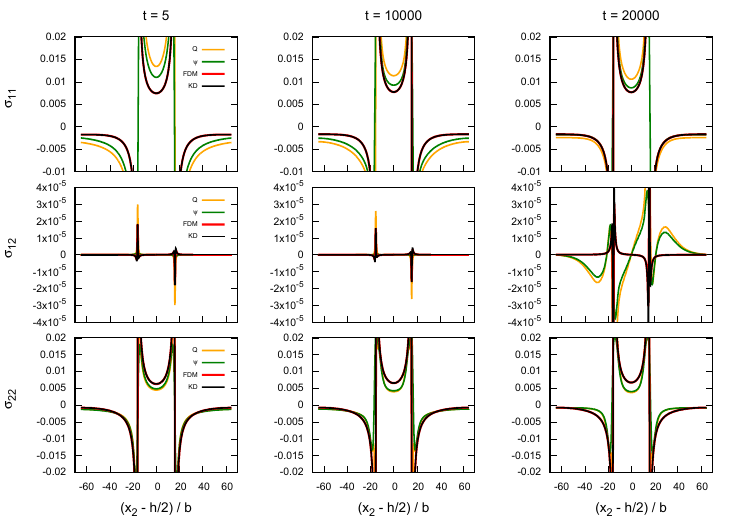}
\caption{Line plots of different stress components as a function of the shifted (with respect to the center of the domain along $x_2$) and normalized (with respect to the Burgers vector magnitude $b$ that is equal to lattice constant $a_0$) distance along $x_2$ passing through the two dislocations at times $t=50$, $t=10000$ and $t=20000$. Stress measures associated with KD correspond to $\bm{\sigma}'$.}\label{fig:stress_vertical}
\end{figure*}

\begin{figure*}[ht!]%
\centering
\includegraphics[width=\textwidth]{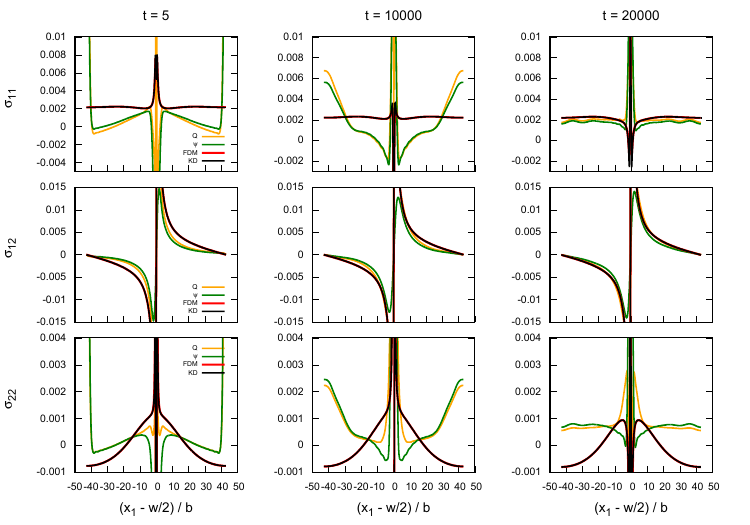}
\caption{Line plots of different stress components as a function of the shifted (with respect to the center of the domain along $x_1$) and normalized (with respect to the Burgers vector magnitude $b$ that is equal to lattice constant $a_0$) distance along $x_1$ passing through the top dislocation at times $t=50$, $10000$ and $20000$. Stress measures associated with KD correspond to $\bm{\sigma}'$.}\label{fig:stress_horizontal}
\end{figure*}

Next, a comparison is performed between the static FDM and PFC stress measures. Note that following Algorithm \ref{alg}, and in order to facilitate comparison between the individual components of all the stress measures, their values have been normalized in such a way that their average over the computational domain is equal to zero. In other words, their zeroth wavenumber ($\mathbf{k} = 0$) component in Fourier space have been set to 0.
We show comparisons of elastic and configurational stresses along two orthogonal lines in the system in Figs. \ref{fig:stress_vertical} and \ref{fig:stress_horizontal}. The various quantities plotted are along the vertical and horizontal lines shown in Figure \ref{fig:figure1}, passing through the two dislocations and the top dislocation, respectively. Analogous plots comparing $\bm{U}$ and $\bm{Q}$ lead to the same conclusions and are not been shown here.

Both figures show that the stress fields $\bm{\sigma}$ and $\bm{\sigma}'$ are identical to numerical accuracy, as expected. 
This agreement serves as validation of the numerical algorithm and the numerical scheme based on the dislocation density as a source (input) in static FDM. 

Figure \ref{fig:stress_vertical} shows all the components of all the stress measures along $x_2$. 
Along this line, the only non negligible components are the normal stress components; the shear stress is very small, and we do not attach any significance to any discrepancies in Figure \ref{fig:stress_vertical}.
At early times ($t=5$), $\sigma^\psi_{11}$ and $\sigma^Q_{11}$ do not agree with $\sigma_{11}$ anywhere along the line. 
The components $\sigma_{22}^Q$ and $\sigma_{22}^\psi$ show better agreement with $\sigma_{22}$. 
Of course, the components of $\bm{\sigma}$ do not change significantly after $t=5$ (negligible changes occur due to slight movement of the dislocation cores), while the configurational stresses continue to relax. 
At long times ($t = 20000$ and beyond), the system has reached a steady state. 
Far field configurational and elastic stresses coincide, however, they do not coincide in the region in between the dislocations. 
This near-field discrepancy is to be expected for several reasons. 
First, the core size in dimensionless units is on the order of $7 \Delta y$ (Fig. \ref{fig:figure1}) which is not much smaller than the distance between dislocations ($144 \Delta y$). 
Second, the phase field free energy is fully nonlinear, while the elastic stress has been obtained under the assumption of linear elasticity. 
Regions of large distortion will therefore have different stress distributions at equilibrium. 
Nevertheless, all stresses coincide far from the defect core region, thus validating the use of the PFC method to provide physically correct equilibrium values of elastic fields.

Similar conclusions can be drawn from an analysis of the stresses $\bm{\sigma}$, $\bm{\sigma^Q}$ and $\bm{\sigma}^\psi$ along $x_1$ (shown in Fig. \ref{fig:stress_horizontal}).
The shear stress $\sigma_{12}$ is not negligible along this direction, with the figure showing good agreement between configurational and elastic stresses. 
There is some disagreement at early times (see $t=5$) because the elastic field is in equilibrium after the dislocation core has stabilized, but the phase field requires a much longer time to relax since it is doing so diffusively. 
At later times, however, the agreement between all the stresses is excellent. 
In addition to the origin of any discrepancies indicated above, the results along this direction are further affected by tensile strains at the edges of the domain; $\sigma^Q_{11}$ and $\sigma^Q_{22}$ deviate significantly from the far field equilibrium solution $\sigma_{11}$ at all times.
This is a consequence of the choice of periodic boundary conditions in a system with a small number of defects, and on the initial condition for the order parameter chosen to include two dislocations.
The arctangent operator in equation (\ref{eqn:psiinit}) is clearly not compatible with periodic boundary conditions. 
Both effects would not be so prominent in a more complex calculation involving an ensemble of defects. In addition, since the values at $\mathbf{k} = 0$ of the various stress fields have been set to zero, their magnitudes in the core region affect the large distance asymptotic values when enforcing this constraint. 
This problem manifests in the results along $x_1$ but it does not manifest in the results along $x_{2}$ due to the different symmetry of the stress field. 

\section{Discussion and conclusions}\label{sec:discussion}

A (weakly) coupled model involving the PFC (Phase Field Crystal) and the static FDM (Field Dislocation Mechanics) approaches has been introduced to obtain the constrained equilibrium state of a prototypical configuration involving a dislocation dipole. 
This configuration allows a critical analysis of the relative role played by configurational (from PFC) and elastic (from static FDM) fields in the coupled theory, the study of the consequences of introducing a consistent separation between diffusive and elastic time scales, and consideration of singularity regularization at defect cores.

The ulterior goal of this study is to endow the PFC model with material elasticity in a way that consistently respects the required independence between the order parameter and lattice distortion. 
We have adopted the view that the PFC order parameter only describes the topological content of the lattice, and that therefore, in equilibrium, it only determines the transverse/incompatible part of the distortion (as required by lattice incompatibility in the elastic problem). 
This choice leverages the fact that the PFC produces a compact and stable defect core, free of any singularity. 
We then use the resulting (diffuse) dislocation density in the core region as the source for the elastic problem.

Our results show that local relaxation of the order parameter at the defect core is fast, and therefore elastic equilibrium should be established rather quickly over the entire system, as shown by the resolution of the elastostatic problem.
However, the configurational stresses have not been equilibrated in the early stages of the numerical calculation. 
It takes a long time for them to reach equilibrium because the phase field order parameter relaxes diffusively everywhere. 
This diffusive relaxation is unphysical for an elastic medium, and it has hindered the applicability of the PFC model to realistic elastic materials.

At much later times, the configurational stresses converge to the elasto-static solution at far field, indicating attainment of equilibrium. 
Any discrepancies remaining are due to the finite size of the computational domain and periodic boundary conditions. 
Near dislocation cores, deviations are found between the elastic and configurational stresses, which are related to the fact that the PFC model is defined by a nonlinear free energy, whereas linear elasticity has been assumed for the static FDM response. 

In summary, the PFC provides a consistent method to introduce lattice incompatibility into a classical elasticity calculation in a defected medium. 
It remains to be investigated whether the same coupled theory provides a proper and efficient description of plastic defect motion as described in \cite{re:acharya20}. 
The algorithm presented in this work will provide the starting point in such an analysis. In addition to the proper separation of order parmameter and elastic time scales, attention has to be paid to a proper separation of dissipative contributions to dislocation motion of phase field and mechanical origin.

\backmatter

\bmhead{Supplementary information}

\bmhead{Acknowledgments}
We are indebted to Amit Acharya for many useful and stimulating discussions. MVU is grateful to the European Research Council (ERC) for their support through the European Union’s Horizon 2020
research and innovation program for project GAMMA (Grant agreement No. 946959). The research of JV has been supported by the National Science Foundation under contract DMR-2223707. 

\section*{Declarations}

The authors declare no conflict of interest associated with this work.



\begin{appendices}

\section{Definitions and notation}\label{sec:notation}

Scalars are denoted with an italic font (e.g., $r$ or $\theta$). Vectors are denoted by a lowercase bold and italic font (e.g., $\boldsymbol{q}$). Unit vectors are identified by an additional overhead hat symbol e.g., $\hat{\bm{e}}$. Second-order tensors are denoted by an uppercase bold and italic font or by bold and italic Greek letters (e.g., $\boldsymbol{U}$ or $\boldsymbol{\alpha}$). The second-order identity tensor is written as $\mathbb{I}$, whose components are $\delta_{ij}$ (the  Kronecker delta). The third-order Levi-Civita permutation tensor is denoted by $\mathbf{X}$, with components $e_{ijk}$ (the permutation symbol). Fourth-order tensors are denoted by double-stroke letters (e.g., $\mathbb{C}$). The null tensor is denoted $0$ for any tensor order. Consider a scalar $\psi$, two vectors $\boldsymbol{u}$ and $\boldsymbol{v}$, two second-order tensors $\boldsymbol{\alpha}$ and $\boldsymbol{B}$, a fourth order tensor $\mathbb{C}$ and a fixed 2-dimensional (2D) Cartesian reference frame with orthonormal basis $\{\hat{\boldsymbol{e}}_i\}$, $i = 1 - d$ ($d=2$) in the $x$ and $y$ directions, respectively. Adopting Einstein notation, the following operations are used in this work:

\vspace{.2cm}

Tensor product: $\boldsymbol{u} \otimes \boldsymbol{v} = u_i v_j \, \hat{\boldsymbol{e}}_i \otimes \hat{\boldsymbol{e}}_j$

\vspace{.2cm}
Inner product: $\begin{aligned}[t]
        &\boldsymbol{u} \cdot \boldsymbol{v} = u_i v_i \\
        &\boldsymbol{\alpha} : \boldsymbol{B} = \alpha_{ij} B_{ij} 
\end{aligned}$

\vspace{.2cm}
Cross product: $\begin{aligned}[t]
    &\boldsymbol{u} \times \boldsymbol{v} = e_{ijk} u_j v_k \, \hat{\boldsymbol{e}}_i \\
    &\boldsymbol{\alpha} \times \boldsymbol{u} = e_{jkl} \alpha_{ik} u_l \, \hat{\boldsymbol{e}}_i \otimes \hat{\boldsymbol{e}}_j
\end{aligned}$

\vspace{.2cm}
Dot and double-dot product: $\begin{aligned}[t]
    &\boldsymbol{\alpha} \cdot \boldsymbol{B} = \alpha_{ij} B_{jk} \, \hat{\boldsymbol{e}}_i \otimes \hat{\boldsymbol{e}}_k \\
    &\boldsymbol{\alpha} \cdot \boldsymbol{u} = \alpha_{ij} u_{j} \, \hat{\boldsymbol{e}}_i \\
    &\boldsymbol{v} \cdot \boldsymbol{B} = v_{i} B_{ij} \, \hat{\boldsymbol{e}}_j \\
    &\mathbf{X} : \boldsymbol{\alpha} = e_{ijk} \alpha_{jk} \hat{\boldsymbol{e}}_i \\
    &\mathbb{C} : \boldsymbol{\alpha} = {C}_{ijkl} \alpha_{kl} \, \hat{\boldsymbol{e}}_i \otimes \hat{\boldsymbol{e}}_j
\end{aligned}$

\vspace{.2cm}
Differential operators: $\begin{aligned}[t]
    &\nabla \psi = \text{grad}\, \psi = \psi_{i}\, \hat{\boldsymbol{e}}_i  \\
    &\nabla \boldsymbol{u} = \text{grad}\, \boldsymbol{u} = u_{i,j}\, \hat{\boldsymbol{e}}_i \otimes \hat{\boldsymbol{e}}_j \\
    &\nabla \boldsymbol{\alpha} = \text{grad}\, \boldsymbol{\alpha} = \alpha_{ij,k}\, \hat{\boldsymbol{e}}_i \otimes \hat{\boldsymbol{e}}_j \otimes \hat{\boldsymbol{e}}_k \\
    &\nabla \cdot \boldsymbol{u} = \text{div}\, \boldsymbol{u} = u_{i,i}\\
    &\nabla \cdot \boldsymbol{\alpha} = \text{div}\, \boldsymbol{\alpha} = \alpha_{ij,j} \, \hat{\boldsymbol{e}}_i\\
    &\nabla \times \boldsymbol{u} = \text{curl}\, \boldsymbol{u} = e_{ijk} u_{k,j}\, \hat{\boldsymbol{e}}_i \\
    &\nabla \times \boldsymbol{\alpha} = \text{curl}\, \boldsymbol{\alpha} = e_{jkl} \alpha_{il,k}\, \hat{\boldsymbol{e}}_i \otimes \hat{\boldsymbol{e}}_j \\
    &\boldsymbol{\alpha} \times \nabla = \text{curl}^\text{T}\, \boldsymbol{\alpha} = e_{ikl} \alpha_{lj,k}\, \hat{\boldsymbol{e}}_i \otimes \hat{\boldsymbol{e}}_j \\
    & \nabla^2 \psi = \nabla \cdot \nabla \psi = \psi_{,ii} \, \\
    & \nabla^2 \bm{\alpha} = \nabla \cdot \nabla \bm{\alpha} = \alpha_{ij,kk} \, \hat{\boldsymbol{e}}_i \otimes \hat{\boldsymbol{e}}_j \, \\
    &\nabla \nabla \psi = \psi_{,ij}  \, \hat{\boldsymbol{e}}_i \otimes \hat{\boldsymbol{e}}_j \,
\end{aligned}$

\vspace{.2cm}
\noindent where the comma indicates differentiation with respect to a given coordinate. Time derivatives are indicated by a superposed dot. 

Overhead tilde signifies a variable in Fourier space. $\mathcal{F}$ is the Fourier transform operator, $\mathcal{F}^{-1}$ is the inverse Fourier transform operator and $\bm{k}$ is the angular wavevector in the Fourier space.

\end{appendices}

\bibliography{sn-bibliography}

\begin{thebibliography}{10}
\expandafter\ifx\csname url\endcsname\relax
  \def\url#1{\burl{#1}}\fi
\expandafter\ifx\csname urlprefix\endcsname\relax\def\urlprefix{URL }\fi
\providecommand{\bibinfo}[2]{#2}
\providecommand{\eprint}[2][]{\url{#2}}
\providecommand{\doi}[1]{\url{https://doi.org/#1}}
\bibcommenthead

\bibitem{Ulvestad2015}
\bibinfo{author}{Ulvestad, A.} \emph{et~al.}
\newblock \bibinfo{title}{Topological defect dynamics in operando battery
  nanoparticles}.
\newblock \emph{\bibinfo{journal}{Science}} \textbf{\bibinfo{volume}{348}},
  \bibinfo{pages}{1344--1347} (\bibinfo{year}{2015}).

\bibitem{Yau2017}
\bibinfo{author}{Yau, A.}, \bibinfo{author}{Cha, W.}, \bibinfo{author}{Kanan,
  M.}, \bibinfo{author}{Stephenson, G.} \& \bibinfo{author}{Ulvestad, A.}
\newblock \bibinfo{title}{Bragg coherent diffractive imaging of single-grain
  defect dynamics in polycrystalline films}.
\newblock \emph{\bibinfo{journal}{Science}} \textbf{\bibinfo{volume}{356}},
  \bibinfo{pages}{739--742} (\bibinfo{year}{2017}).

\bibitem{re:shen20}
\bibinfo{author}{Shen, Y.-F.}, \bibinfo{author}{Liu, H.} \&
  \bibinfo{author}{Suter, R.~M.}
\newblock \bibinfo{title}{Voxel-based strain tensors from near-field high
  energy diffraction microscopy}.
\newblock \emph{\bibinfo{journal}{Current Opinion in Solid State and Materials
  Science}} \textbf{\bibinfo{volume}{24}}, \bibinfo{pages}{100852}
  (\bibinfo{year}{2020}).

\bibitem{re:wilkin21}
\bibinfo{author}{Wilkin, M.~J.} \emph{et~al.}
\newblock \bibinfo{title}{Experimental demonstration of coupled multi-peak
  bragg coherent diffraction imaging with genetic algorithms}.
\newblock \emph{\bibinfo{journal}{Physical Review B}}
  \textbf{\bibinfo{volume}{103}} (\bibinfo{year}{2021}).

\bibitem{re:dresselhaus21}
\bibinfo{author}{Dresselhaus-Marais, L.~E.} \emph{et~al.}
\newblock \bibinfo{title}{In situ visualization of long-range defect
  interactions at the edge of melting}.
\newblock \emph{\bibinfo{journal}{Science Advances}}
  \textbf{\bibinfo{volume}{7}}, \bibinfo{pages}{eabe8311}
  (\bibinfo{year}{2021}).

\bibitem{re:suter17}
\bibinfo{author}{Suter, R.}
\newblock \bibinfo{title}{Multiscale measurements for materials modeling}.
\newblock \emph{\bibinfo{journal}{Science}} \textbf{\bibinfo{volume}{356}},
  \bibinfo{pages}{704--705} (\bibinfo{year}{2017}).
\newblock \urlprefix\url{http://science.sciencemag.org/content/356/6339/704}.

\bibitem{Hanson2018}
\bibinfo{author}{Hanson, J.} \emph{et~al.}
\newblock \bibinfo{title}{Crystallographic character of grain boundaries
  resistant to hydrogen-assisted fracture in ni-basealloy 725}.
\newblock \emph{\bibinfo{journal}{Nature Communications}}
  \textbf{\bibinfo{volume}{9}}, \bibinfo{pages}{3386} (\bibinfo{year}{2018}).

\bibitem{Bhattacharya2019}
\bibinfo{author}{Bhattacharya, A.} \emph{et~al.}
\newblock \bibinfo{title}{Three-dimensional observations of grain volume
  changes during annealing of polycrystalline ni}.
\newblock \emph{\bibinfo{journal}{Acta Materialia}}
  \textbf{\bibinfo{volume}{167}}, \bibinfo{pages}{40--50}
  (\bibinfo{year}{2019}).

\bibitem{acharya_model_2001}
\bibinfo{author}{Acharya, A.}
\newblock \bibinfo{title}{A model of crystal plasticity based on the theory of
  continuously distributed dislocations} \textbf{\bibinfo{volume}{49}},
  \bibinfo{pages}{761--784}.
\newblock
  \urlprefix\url{https://www.sciencedirect.com/science/article/pii/S0022509600000600}.

\bibitem{acharya_driving_2003}
\bibinfo{author}{Acharya, A.}
\newblock \bibinfo{title}{Driving forces and boundary conditions in continuum
  dislocation mechanics} \textbf{\bibinfo{volume}{459}},
  \bibinfo{pages}{1343--1363}.
\newblock
  \urlprefix\url{https://royalsocietypublishing.org/doi/abs/10.1098/rspa.2002.1095}.

\bibitem{acharya_constitutive_2004}
\bibinfo{author}{Acharya, A.}
\newblock \bibinfo{title}{Constitutive analysis of finite deformation field
  dislocation mechanics} \textbf{\bibinfo{volume}{52}},
  \bibinfo{pages}{301--316}.
\newblock
  \urlprefix\url{http://www.sciencedirect.com/science/article/pii/S0022509603000930}.

\bibitem{roy2005finite}
\bibinfo{author}{Roy, A.} \& \bibinfo{author}{Acharya, A.}
\newblock \bibinfo{title}{Finite element approximation of field dislocation
  mechanics}.
\newblock \emph{\bibinfo{journal}{Journal of the Mechanics and Physics of
  Solids}} \textbf{\bibinfo{volume}{53}}, \bibinfo{pages}{143--170}
  (\bibinfo{year}{2005}).

\bibitem{acharya_size_2006}
\bibinfo{author}{Acharya, A.} \& \bibinfo{author}{Roy, A.}
\newblock \bibinfo{title}{Size effects and idealized dislocation microstructure
  at small scales: Predictions of a phenomenological model of mesoscopic field
  dislocation mechanics: Part i} \textbf{\bibinfo{volume}{54}},
  \bibinfo{pages}{1687--1710}.

\bibitem{re:acharya11}
\bibinfo{author}{Acharya, A.}
\newblock \bibinfo{title}{Microcanonical entropy and mesoscale dislocation
  mechanics and plasticity}.
\newblock \emph{\bibinfo{journal}{Journal of Elasticity}}
  \textbf{\bibinfo{volume}{104}}, \bibinfo{pages}{23--44}
  (\bibinfo{year}{2011}).

\bibitem{zhang2015single}
\bibinfo{author}{Zhang, X.}, \bibinfo{author}{Acharya, A.},
  \bibinfo{author}{Walkington, N.~J.} \& \bibinfo{author}{Bielak, J.}
\newblock \bibinfo{title}{A single theory for some quasi-static, supersonic,
  atomic, and tectonic scale applications of dislocations}.
\newblock \emph{\bibinfo{journal}{Journal of the Mechanics and Physics of
  Solids}} \textbf{\bibinfo{volume}{84}}, \bibinfo{pages}{145--195}
  (\bibinfo{year}{2015}).

\bibitem{arora2020finite}
\bibinfo{author}{Arora, R.}, \bibinfo{author}{Zhang, X.} \&
  \bibinfo{author}{Acharya, A.}
\newblock \bibinfo{title}{Finite element approximation of finite deformation
  dislocation mechanics}.
\newblock \emph{\bibinfo{journal}{Computer Methods in Applied Mechanics and
  Engineering}} \textbf{\bibinfo{volume}{367}}, \bibinfo{pages}{113076}
  (\bibinfo{year}{2020}).

\bibitem{re:arora23}
\bibinfo{author}{Arora, A.}, \bibinfo{author}{Arora, R.} \&
  \bibinfo{author}{Acharya, A.}
\newblock \bibinfo{title}{Interface-dominated plasticity and kink bands in
  metallic nanolaminates}.
\newblock \emph{\bibinfo{journal}{Crystals}} \textbf{\bibinfo{volume}{13}},
  \bibinfo{pages}{828} (\bibinfo{year}{2023}).

\bibitem{re:haataja02}
\bibinfo{author}{Haataja, M.}, \bibinfo{author}{{M\"uller}, J.},
  \bibinfo{author}{Rutenberg, A.~D.} \& \bibinfo{author}{Grant, M.}
\newblock \bibinfo{title}{Dislocations and morphological instabilities:
  Continuum modeling of misfitting heteroepitaxial films}.
\newblock \emph{\bibinfo{journal}{Phys. Rev. B}} \textbf{\bibinfo{volume}{65}},
  \bibinfo{pages}{165414} (\bibinfo{year}{2002}).

\bibitem{re:elder02}
\bibinfo{author}{Elder, K.}, \bibinfo{author}{Katakowski, M.},
  \bibinfo{author}{Haataja, M.} \& \bibinfo{author}{Grant, M.}
\newblock \bibinfo{title}{Modeling elasticity in crystal growth}.
\newblock \emph{\bibinfo{journal}{Phys. Rev. Lett.}}
  \textbf{\bibinfo{volume}{88}}, \bibinfo{pages}{245701}
  (\bibinfo{year}{2002}).

\bibitem{re:elder04}
\bibinfo{author}{Elder, K.~R.} \& \bibinfo{author}{Grant, M.}
\newblock \bibinfo{title}{Modeling elastic and plastic deformations in
  nonequilibrium processing using phase field crystals}.
\newblock \emph{\bibinfo{journal}{Phys. Rev. E}} \textbf{\bibinfo{volume}{70}},
  \bibinfo{pages}{051605} (\bibinfo{year}{2004}).
\newblock \urlprefix\url{https://link.aps.org/doi/10.1103/PhysRevE.70.051605}.

\bibitem{re:salvalaglio19}
\bibinfo{author}{Salvalaglio, M.}, \bibinfo{author}{Voigt, A.} \&
  \bibinfo{author}{Elder, K.~R.}
\newblock \bibinfo{title}{Closing the gap between atomic-scale lattice
  deformations and continuum elasticity}.
\newblock \emph{\bibinfo{journal}{npj Computational Materials}}
  \textbf{\bibinfo{volume}{5}}, \bibinfo{pages}{48} (\bibinfo{year}{2019}).

\bibitem{re:zhao23}
\bibinfo{author}{Zhao, Y.}
\newblock \bibinfo{title}{Understanding and design of metallic alloys guided by
  phase-field simulations}.
\newblock \emph{\bibinfo{journal}{npj Computational Materials}}
  \textbf{\bibinfo{volume}{9}}, \bibinfo{pages}{94} (\bibinfo{year}{2023}).

\bibitem{re:skogvoll23}
\bibinfo{author}{Skogvoll, V.}, \bibinfo{author}{R{\o}nning, J.},
  \bibinfo{author}{Salvalaglio, M.} \& \bibinfo{author}{Angheluta, L.}
\newblock \bibinfo{title}{A unified field theory of topological defects and
  non-linear local excitations}.
\newblock \emph{\bibinfo{journal}{npj Computational Materials}}
  \textbf{\bibinfo{volume}{9}}, \bibinfo{pages}{122} (\bibinfo{year}{2023}).

\bibitem{re:emmerich12}
\bibinfo{author}{Emmerich, H.} \emph{et~al.}
\newblock \bibinfo{title}{Phase-field-crystal models for condensed matter
  dynamics on atomic length and diffusive time scales: an overview}.
\newblock \emph{\bibinfo{journal}{Adv. in Physics}}
  \textbf{\bibinfo{volume}{61}}, \bibinfo{pages}{665--743}
  (\bibinfo{year}{2012}).

\bibitem{re:momeni18}
\bibinfo{author}{Momeni, K.}, \bibinfo{author}{Ji, Y.}, \bibinfo{author}{Zhang,
  K.}, \bibinfo{author}{Robinson, J.~A.} \& \bibinfo{author}{Chen, L.-Q.}
\newblock \bibinfo{title}{Multiscale framework for simulation-guided growth of
  2d materials}.
\newblock \emph{\bibinfo{journal}{npj 2D Materials and Applications}}
  \textbf{\bibinfo{volume}{2}}, \bibinfo{pages}{1--7} (\bibinfo{year}{2018}).

\bibitem{zhang2017continuum}
\bibinfo{author}{Zhang, X.}
\newblock \bibinfo{title}{A continuum model for dislocation pile-up problems}.
\newblock \emph{\bibinfo{journal}{Acta Materialia}}
  \textbf{\bibinfo{volume}{128}}, \bibinfo{pages}{428--439}
  (\bibinfo{year}{2017}).

\bibitem{re:arora20a}
\bibinfo{author}{Arora, R.}, \bibinfo{author}{Zhang, X.} \&
  \bibinfo{author}{Acharya, A.}
\newblock \bibinfo{title}{Finite element approximation of finite deformation
  dislocation mechanics}.
\newblock \emph{\bibinfo{journal}{Computer Methods in Applied Mechanics and
  Engineering}} \textbf{\bibinfo{volume}{367}}, \bibinfo{pages}{113076}
  (\bibinfo{year}{2020}).

\bibitem{re:arora20b}
\bibinfo{author}{Arora, R.} \& \bibinfo{author}{Acharya, A.}
\newblock \bibinfo{title}{A unification of finite deformation ${J}_2$
  von-{M}ises plasticity and quantitative dislocation mechanics}.
\newblock \emph{\bibinfo{journal}{Journal of the Mechanics and Physics of
  Solids}} \textbf{\bibinfo{volume}{143}}, \bibinfo{pages}{104050}
  (\bibinfo{year}{2020}).

\bibitem{re:kosevich79}
\bibinfo{author}{Kosevich, A.}
\newblock \bibinfo{title}{ in \textit{Crystal dislocations and the theory of
  elasticity}} (ed.\bibinfo{editor}{Nabarro, F.})
  \emph{\bibinfo{booktitle}{Dislocations in Solids}}, Vol.~\bibinfo{volume}{1}
  \bibinfo{pages}{33} (\bibinfo{publisher}{North-Holland},
  \bibinfo{address}{New York}, \bibinfo{year}{1979}).

\bibitem{re:rickman97}
\bibinfo{author}{Rickman, J.} \& \bibinfo{author}{{Vi\~nals}, J.}
\newblock \bibinfo{title}{Modeling of dislocation structures in materials}.
\newblock \emph{\bibinfo{journal}{Phil. Mag. A}} \textbf{\bibinfo{volume}{75}},
  \bibinfo{pages}{1251} (\bibinfo{year}{1997}).

\bibitem{re:perreault16}
\bibinfo{author}{Perreault, B.}, \bibinfo{author}{Vinals, J.} \&
  \bibinfo{author}{Rickman, J.~M.}
\newblock \bibinfo{title}{Impact of lattice rotation on dislocation motion}.
\newblock \emph{\bibinfo{journal}{Physical Review B}}
  \textbf{\bibinfo{volume}{93}}, \bibinfo{pages}{014107}
  (\bibinfo{year}{2016}).

\bibitem{UPADHYAY2020104150}
\bibinfo{author}{Upadhyay, M.~V.}
\newblock \bibinfo{title}{On the thermo-mechanical theory of field dislocations
  in transient heterogeneous temperature fields}.
\newblock \emph{\bibinfo{journal}{Journal of the Mechanics and Physics of
  Solids}} \textbf{\bibinfo{volume}{145}}, \bibinfo{pages}{104150}
  (\bibinfo{year}{2020}).
\newblock
  \urlprefix\url{https://www.sciencedirect.com/science/article/pii/S0022509620303835}.

\bibitem{LIMACHAVES2024116763}
\bibinfo{author}{Lima-Chaves, G.~D.} \& \bibinfo{author}{Upadhyay, M.~V.}
\newblock \bibinfo{title}{Finite element implementation of the thermal field
  dislocation mechanics model: Study of temperature evolution due to
  dislocation activity}.
\newblock \emph{\bibinfo{journal}{Computer Methods in Applied Mechanics and
  Engineering}} \textbf{\bibinfo{volume}{421}}, \bibinfo{pages}{116763}
  (\bibinfo{year}{2024}).
\newblock
  \urlprefix\url{https://www.sciencedirect.com/science/article/pii/S0045782524000197}.

\bibitem{vitek1968intrinsic}
\bibinfo{author}{Vitek, V.}
\newblock \bibinfo{title}{Intrinsic stacking faults in body-centred cubic
  crystals}.
\newblock \emph{\bibinfo{journal}{Philosophical Magazine}}
  \textbf{\bibinfo{volume}{18}}, \bibinfo{pages}{773--786}
  (\bibinfo{year}{1968}).

\bibitem{vitek1998atomic}
\bibinfo{author}{Vitek, V.}
\newblock \bibinfo{title}{Atomic structure of dislocations in intermetallics
  with close packed structures: a comparative study}.
\newblock \emph{\bibinfo{journal}{Intermetallics}}
  \textbf{\bibinfo{volume}{6}}, \bibinfo{pages}{579--585}
  (\bibinfo{year}{1998}).

\bibitem{vitek2008non}
\bibinfo{author}{Vitek, V.} \& \bibinfo{author}{Paidar, V.}
\newblock \bibinfo{title}{Non-planar dislocation cores: a ubiquitous phenomenon
  affecting mechanical properties of crystalline materials}.
\newblock \emph{\bibinfo{journal}{Dislocations in solids}}
  \textbf{\bibinfo{volume}{14}}, \bibinfo{pages}{439--514}
  (\bibinfo{year}{2008}).

\bibitem{re:elder07}
\bibinfo{author}{Elder, K.~R.}, \bibinfo{author}{Provatas, N.},
  \bibinfo{author}{Berry, J.}, \bibinfo{author}{Stefanovic, P.} \&
  \bibinfo{author}{Grant, M.}
\newblock \bibinfo{title}{Phase-field crystal modeling and classical density
  functional theory of freezing}.
\newblock \emph{\bibinfo{journal}{Phys. Rev. B}} \textbf{\bibinfo{volume}{75}},
  \bibinfo{pages}{064107} (\bibinfo{year}{2007}).

\bibitem{re:berry06}
\bibinfo{author}{Berry, J.}, \bibinfo{author}{Grant, M.} \&
  \bibinfo{author}{Elder, K.}
\newblock \bibinfo{title}{Diffusive atomistic dynamics of edge dislocations in
  two dimensions}.
\newblock \emph{\bibinfo{journal}{Phys. Rev. E}} \textbf{\bibinfo{volume}{73}},
  \bibinfo{pages}{031609} (\bibinfo{year}{2006}).

\bibitem{re:skaugen18}
\bibinfo{author}{Skaugen, A.}, \bibinfo{author}{Angheluta, L.} \&
  \bibinfo{author}{Vi{\~n}als, J.}
\newblock \bibinfo{title}{Dislocation dynamics and crystal plasticity in the
  phase-field crystal model}.
\newblock \emph{\bibinfo{journal}{Phys. Rev. B}} \textbf{\bibinfo{volume}{97}},
  \bibinfo{pages}{054113} (\bibinfo{year}{2018}).

\bibitem{re:skaugen18b}
\bibinfo{author}{Skaugen, A.}, \bibinfo{author}{Angheluta, L.} \&
  \bibinfo{author}{Vi{\~n}als, J.}
\newblock \bibinfo{title}{Separation of elastic and plastic timescales in a
  phase field crystal model}.
\newblock \emph{\bibinfo{journal}{Phys. Rev. Lett.}}
  \textbf{\bibinfo{volume}{121}}, \bibinfo{pages}{255501}
  (\bibinfo{year}{2018}).

\bibitem{re:skogvoll22}
\bibinfo{author}{Skogvoll, V.}, \bibinfo{author}{Angheluta, L.},
  \bibinfo{author}{Skaugen, A.}, \bibinfo{author}{Salvalaglio, M.} \&
  \bibinfo{author}{Vi{\~n}als, J.}
\newblock \bibinfo{title}{A phase field crystal theory of the kinematics of
  dislocation lines}.
\newblock \emph{\bibinfo{journal}{Journal of the Mechanics and Physics of
  Solids}} \bibinfo{pages}{104932} (\bibinfo{year}{2022}).

\bibitem{re:mellenthin08}
\bibinfo{author}{Mellenthin, J.}, \bibinfo{author}{Karma, A.} \&
  \bibinfo{author}{Plapp, M.}
\newblock \bibinfo{title}{Phase-field crystal study of grain-boundary
  premelting}.
\newblock \emph{\bibinfo{journal}{Phys. Revi. B}}
  \textbf{\bibinfo{volume}{78}}, \bibinfo{pages}{184110}
  (\bibinfo{year}{2008}).

\bibitem{re:wu12}
\bibinfo{author}{Wu, K.-A.} \& \bibinfo{author}{Voorhees, P.~W.}
\newblock \bibinfo{title}{Phase field crystal simulations of nanocrystalline
  grain growth in two dimensions}.
\newblock \emph{\bibinfo{journal}{Acta Materialia}}
  \textbf{\bibinfo{volume}{60}}, \bibinfo{pages}{407--419}
  (\bibinfo{year}{2012}).

\bibitem{re:heinonen16}
\bibinfo{author}{Heinonen, V.} \emph{et~al.}
\newblock \bibinfo{title}{Consistent hydrodynamics for phase field crystals}.
\newblock \emph{\bibinfo{journal}{Phys. Rev. Lett.}}
  \textbf{\bibinfo{volume}{116}}, \bibinfo{pages}{024303}
  (\bibinfo{year}{2016}).

\bibitem{re:acharya22}
\bibinfo{author}{Acharya, A.}, \bibinfo{author}{Angheluta, L.} \&
  \bibinfo{author}{{Vi\~nals}, J.}
\newblock \bibinfo{title}{Elasticity versus phase field driven motion in the
  phase field crystal model}.
\newblock \emph{\bibinfo{journal}{Modelling Simul. Mater. Sci. Eng.}}
  \textbf{\bibinfo{volume}{30}}, \bibinfo{pages}{064005}
  (\bibinfo{year}{2022}).

\bibitem{re:mura63}
\bibinfo{author}{Mura, T.}
\newblock \bibinfo{title}{Continuous distribution of moving dislocations}.
\newblock \emph{\bibinfo{journal}{Philosophical Magazine}}
  \textbf{\bibinfo{volume}{8}}, \bibinfo{pages}{843--857}
  (\bibinfo{year}{1963}).

\bibitem{re:acharya20}
\bibinfo{author}{Acharya, A.} \& \bibinfo{author}{{Vi\~nals}, J.}
\newblock \bibinfo{title}{Field dislocation mechanics and phase field crystal
  models}.
\newblock \emph{\bibinfo{journal}{Physical Review B}}
  \textbf{\bibinfo{volume}{102}}, \bibinfo{pages}{064109}
  (\bibinfo{year}{2020}).

\bibitem{elder_phase-field_2007}
\bibinfo{author}{Elder, K.~R.}, \bibinfo{author}{Provatas, N.},
  \bibinfo{author}{Berry, J.}, \bibinfo{author}{Stefanovic, P.} \&
  \bibinfo{author}{Grant, M.}
\newblock \bibinfo{title}{Phase-field crystal modeling and classical density
  functional theory of freezing}.
\newblock \emph{\bibinfo{journal}{Physical Review B}}
  \textbf{\bibinfo{volume}{75}}, \bibinfo{pages}{064107}
  (\bibinfo{year}{2007}).
\newblock \urlprefix\url{https://link.aps.org/doi/10.1103/PhysRevB.75.064107}.

\bibitem{re:cross93}
\bibinfo{author}{Cross, M.~C.} \& \bibinfo{author}{Hohenberg, P.~C.}
\newblock \bibinfo{title}{Pattern formation outside of equilibrium}.
\newblock \emph{\bibinfo{journal}{Rev. Mod. Phys.}}
  \textbf{\bibinfo{volume}{65}}, \bibinfo{pages}{851--1112}
  (\bibinfo{year}{1993}).
\newblock \urlprefix\url{https://link.aps.org/doi/10.1103/RevModPhys.65.851}.

\bibitem{re:siggia81}
\bibinfo{author}{Siggia, E.~D.} \& \bibinfo{author}{Zippelius, A.}
\newblock \bibinfo{title}{Dynamics of defects in rayleigh-b\'enard convection}.
\newblock \emph{\bibinfo{journal}{Phys. Rev. A}} \textbf{\bibinfo{volume}{24}},
  \bibinfo{pages}{1036--1049} (\bibinfo{year}{1981}).
\newblock \urlprefix\url{https://link.aps.org/doi/10.1103/PhysRevA.24.1036}.

\bibitem{re:tsimring96}
\bibinfo{author}{Tsimring, L.~S.}
\newblock \bibinfo{title}{Dynamics of penta-hepta defects in hexagonal
  patterns}.
\newblock \emph{\bibinfo{journal}{Physica D: Nonlinear Phenomena}}
  \textbf{\bibinfo{volume}{89}}, \bibinfo{pages}{368--380}
  (\bibinfo{year}{1996}).

\bibitem{re:boyer02b}
\bibinfo{author}{Boyer, D.} \& \bibinfo{author}{{Vi\~nals}, J.}
\newblock \bibinfo{title}{Weakly nonlinear theory of grain boundary motion in
  patterns with crystalline symmetry}.
\newblock \emph{\bibinfo{journal}{Phys. Rev. Lett.}}
  \textbf{\bibinfo{volume}{89}}, \bibinfo{pages}{055501}
  (\bibinfo{year}{2002}).

\bibitem{re:dewit60}
\bibinfo{author}{de~Wit, R.}
\newblock \bibinfo{title}{The continuum theory of stationary dislocations}.
\newblock \emph{\bibinfo{journal}{Solid state physics}}
  \textbf{\bibinfo{volume}{10}}, \bibinfo{pages}{249--292}
  (\bibinfo{year}{1960}).

\bibitem{re:kroner81}
\bibinfo{author}{{Kr\"oner}, E.}
\newblock \bibinfo{editor}{Balian, R.}, \bibinfo{editor}{{Kl\'eman}, M.} \&
  \bibinfo{editor}{Poirier, J.} (eds) \emph{\bibinfo{title}{Continuum theory of
  defects}}.
\newblock (eds \bibinfo{editor}{Balian, R.}, \bibinfo{editor}{{Kl\'eman}, M.}
  \& \bibinfo{editor}{Poirier, J.}) \emph{\bibinfo{booktitle}{Physics of
  Defects}} (\bibinfo{publisher}{North Holland}, \bibinfo{year}{1981}).
\newblock \bibinfo{note}{Proceedings of Les Houches, Session XXXV}.

\bibitem{FFTW05}
\bibinfo{author}{Frigo, M.} \& \bibinfo{author}{Johnson, S.~G.}
\newblock \bibinfo{title}{The design and implementation of {FFTW3}}.
\newblock \emph{\bibinfo{journal}{Proceedings of the IEEE}}
  \textbf{\bibinfo{volume}{93}}, \bibinfo{pages}{216--231}
  (\bibinfo{year}{2005}).
\newblock \bibinfo{note}{Special issue on ``Program Generation, Optimization,
  and Platform Adaptation''}.

\bibitem{BERBENNI20144157}
\bibinfo{author}{Berbenni, S.}, \bibinfo{author}{Taupin, V.},
  \bibinfo{author}{Djaka, K.~S.} \& \bibinfo{author}{Fressengeas, C.}
\newblock \bibinfo{title}{A numerical spectral approach for solving
  elasto-static field dislocation and g-disclination mechanics}.
\newblock \emph{\bibinfo{journal}{International Journal of Solids and
  Structures}} \textbf{\bibinfo{volume}{51}}, \bibinfo{pages}{4157--4175}
  (\bibinfo{year}{2014}).
\newblock
  \urlprefix\url{https://www.sciencedirect.com/science/article/pii/S0020768314003138}.

\bibitem{UPADHYAY2016126}
\bibinfo{author}{Upadhyay, M.}, \bibinfo{author}{Capolungo, L.},
  \bibinfo{author}{Taupin, V.}, \bibinfo{author}{Fressengeas, C.} \&
  \bibinfo{author}{Lebensohn, R.}
\newblock \bibinfo{title}{A higher order elasto-viscoplastic model using fast
  fourier transforms: Effects of lattice curvatures on mechanical response of
  nanocrystalline metals}.
\newblock \emph{\bibinfo{journal}{International Journal of Plasticity}}
  \textbf{\bibinfo{volume}{83}}, \bibinfo{pages}{126--152}
  (\bibinfo{year}{2016}).
\newblock
  \urlprefix\url{https://www.sciencedirect.com/science/article/pii/S0749641916300560}.

\end{thebibliography}

\end{document}